\documentclass[runningheads]{llncs}
\usepackage[T1]{fontenc}
\usepackage{graphicx}

\usepackage{hyperref}
\begin{document}
\title{Mitigating the Carbon Footprint of Chatbots\\ as Consumers}

\titlerunning{Mitigating the Carbon Footprint of Chatbots as Consumers}
%
%
\author{Boris Ruf \and
Marcin Detyniecki}
\authorrunning{B. Ruf et al.}

%
\institute{AXA Group Operations, AI Research, Paris, France
\email{\{boris.ruf,marcin.detyniecki\}@axa.com}}
\maketitle              
\begin{abstract}
In the context of the high energy demand of large language models (LLMs) and growing concerns about global warming, there is significant demand for actionable recommendations that can help reduce emissions when utilizing such technologies. This paper examines the environmental impact linked to a fundamental function of LLM-based conversational systems that might be less well known to end users: the conversational memory, which enables the system to maintain context throughout the dialog. After analyzing conversation patterns using anonymized token data from a real world system, a recommendation for individuals on how they could use chatbots in a more sustainable way is derived. Based on a simulation, the savings potential resulting from the adoption of such an ecological gesture is estimated.
\end{abstract}

\section{Motivation}
Artificial intelligence (AI) is widely recognized as a disruptive technology that has the potential to transform various aspects of our lives and revolutionize the way we work. Successes in natural language processing and machine learning have led to the development of large language models (LLMs) such as ChatGPT. These AI-based conversational agents are gaining ground in various fields, from programming to customer service and education, with their ability to engage in natural, human-like conversation~\cite{10.1145/3581641.3584037,kasneci2023chatgpt,thirunavukarasu2023large}. The drawback of such technology is that its development and application consumes enormous amounts of energy and leaves a large carbon footprint~\cite{10.1145/3381831,Luccioni_2024}. This is particularly concerning, as the imminent global warming crisis requires us to significantly reduce carbon emissions to mitigate its impact on society~\cite{IPCC_2022_WGIII}. So, how can we continue to benefit from the innovative potential of AI and, at the same time, reduce its carbon footprint? 

AI providers and data centers hold the greatest potential for optimizing efficiency. End users, in contrast, have limited options and there are few strong recommendations available~\cite{rong2016optimizing}. To narrow this gap, this paper suggests an uncomplicated measure derived from analyzing the structure and operations of conversational agents employing LLMs. 

The remaining parts of this paper are organized as follows: First, relevant sources to related research is provided. Next, some background information on the mechanisms of chatbots that use LLMs is presented. Subsequently, the recommendation for sustainable action is derived. Lastly, its potential impact is assessed in an empirical study.


\section{Related work}

Several research papers have brought attention to the substantial energy consumption associated with AI model training and inference~\cite{rillig2023risks,luccioni2023estimating,luccioni2023power}. The influence of LLM-based conversational agents such as ChatGPT in various domains has been thoroughly examined~\cite{RAY2023121}, and in particular the evolution of human interaction with dialog-based user interface systems has been the subject of extensive research~\cite{rapp2021human,ARAUJO2018183}. Furthermore, behavioral research has explored sustainable consumer behavior and the obstacles preventing its adoption~\cite{zimmermann2021digital,RePEc:eme:srjpps:srj-05-2020-0203}.

\begin{figure}[t]
\centering
\includegraphics[width=0.7\linewidth]{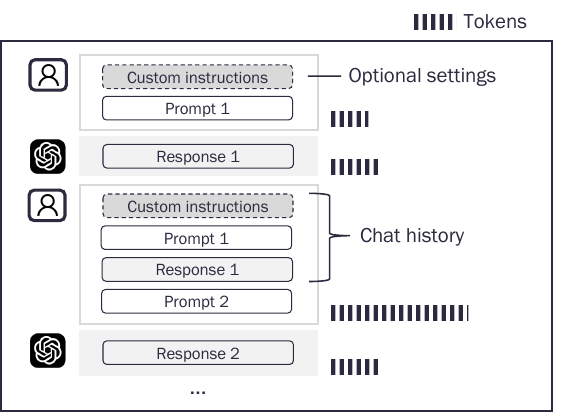}

    \caption{Schematic illustration of how conversational memory works in its most basic version. Notably, the chat history is included in every new prompt.}
    \label{fig:conversational_memory}
  \end{figure}

\section{Background}

In contrast to conventional API calls that mainly involve a database lookup to retrieve precomputed information, interacting with LLMs is very computationally intensive. Those models are typically made up of complex artificial neural networks that have multiple layers, each containing millions of parameters. The query process involves running input data through these network layers. Due to the large number of parameters and computations involved, this operation becomes highly complex and requires a considerable amount of processing power~\cite{NIPS2017_3f5ee243}.

The input data of LLMs are tokenized, where each token represents an individual unit of text such as a word or character~\cite{10.1162/coli_a_00474}. The number of tokens strongly affects the computational load and energy consumption of the model. Billing models are often also token-based.

\begin{figure}[ht]
\centering

\includegraphics[width=0.8\linewidth]{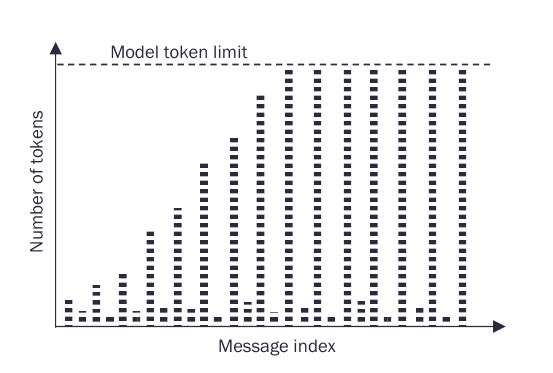}
    \caption{Exemplary token usage during a single conversation. It shows an increase in prompt tokens as the number of messages exchanged rises, while the number of completion tokens remains consistently low.}
    \label{fig:token_usage}
 
\end{figure}

By default, LLMs are stateless and process incoming queries independently. To ensure contextual consistency across conversations and produce human-like responses, conversational agents utilize conversational memory. In its most basic version, previous user prompts but also model responses are stored in its raw form and included as chat history in each new request, as illustrated in simplified form in Figure~\ref{fig:conversational_memory}. Unnoticed by the user, optional custom instructions or templates which define the chatbot's behavior may get included in the first request and  further increase the conversational memory. Altogether, this concept results in an increasing number of tokens in the input data as the conversation progresses until the model's token limit is reached (Figure~\ref{fig:token_usage}). This limit is, for example, 4,096 for gpt-3.5-turbo and Llama2\footnote{https://huggingface.co/docs/transformers/main/model\_doc/llama2}, and 128,000 for gpt-4-1106-preview\footnote{https://platform.openai.com/docs/models/gpt-4-and-gpt-4-turbo}.

To partly mitigate this effect, more sophisticated approaches to create conversational memory are available. Conversation summarization compresses the chat history but requires additional use of tokens~\cite{xu-etal-2022-beyond,zhong-etal-2022-less}. Buffer windows set arbitrary limits for the memory size.\footnote{https://www.pinecone.io/learn/series/langchain/langchain-conversational-memory} However, in principle it can be observed: The conversation becomes more expensive, both in environmental and financial terms, the longer it lasts.

\section{Recommendation}
Drawing from the observations above, the following recommendation is formulated as an effective eco gesture for end users of conversational agents: Reset the conversation whenever the topic or the purpose of the interaction changes and the preceding messages become irrelevant. Without compromising the quality of the answers, such a measure can help reduce the number of tokens used and thus the energy consumption.

\section{Empirical study}
To validate the eco gesture and assess its effectiveness, its adoption is modeled with usage data from a real world system.

\subsection{Dataset}
The dataset used originates from a conversational system based on a GPT-3.5-turbo model. The chatbot is operated internally in an organization; the dataset is not publicly accessible. The anonymized log files consist of 190,120 rows and capture a 30-day operational period. The dataset includes the fields \textit{datetime\_UTC}, \textit{user\_id}, \textit{prompt\_tokens}, and \textit{completion\_tokens}, but it does not encompass message content. The field \textit{completion\_tokens} is disregarded in the current analysis, as it holds no relevance to the object of investigation.

\subsection{Data preparation}
In order to deal with the lack of a distinct identifier for each conversational thread, the following heuristic approach was applied. A helper request, utilized to generate a title for the thread, was found to be a useful signal for the initiation of a new thread: Its timestamp is identical to that of the first user prompt, and its number of prompt tokens exceeds the number of tokens in the initial prompt by a constant value. After these markers were used to label the different conversational threads, the helper requests were deleted from the dataset. To consider the possibility of users engaging in multiple conversations simultaneously, the next step was to remove irregular conversation threads in which the number of prompt tokens does not increase monotonically over time. This process affected 493 threads. After the data cleansing, 144,247 rows remained, representing 40,065 conversation threads. Subsequently, the message index, which indicates the position within the conversation, was added as a supplementary data field for ease of reference. Finally, to determine the pauses, the time gap between each message and the preceding message within the same conversational thread was calculated. This duration was then allocated to the field named \textit{pause\_in\_seconds}.

\begin{figure}[h]
\centering
\includegraphics[width=0.9\linewidth]{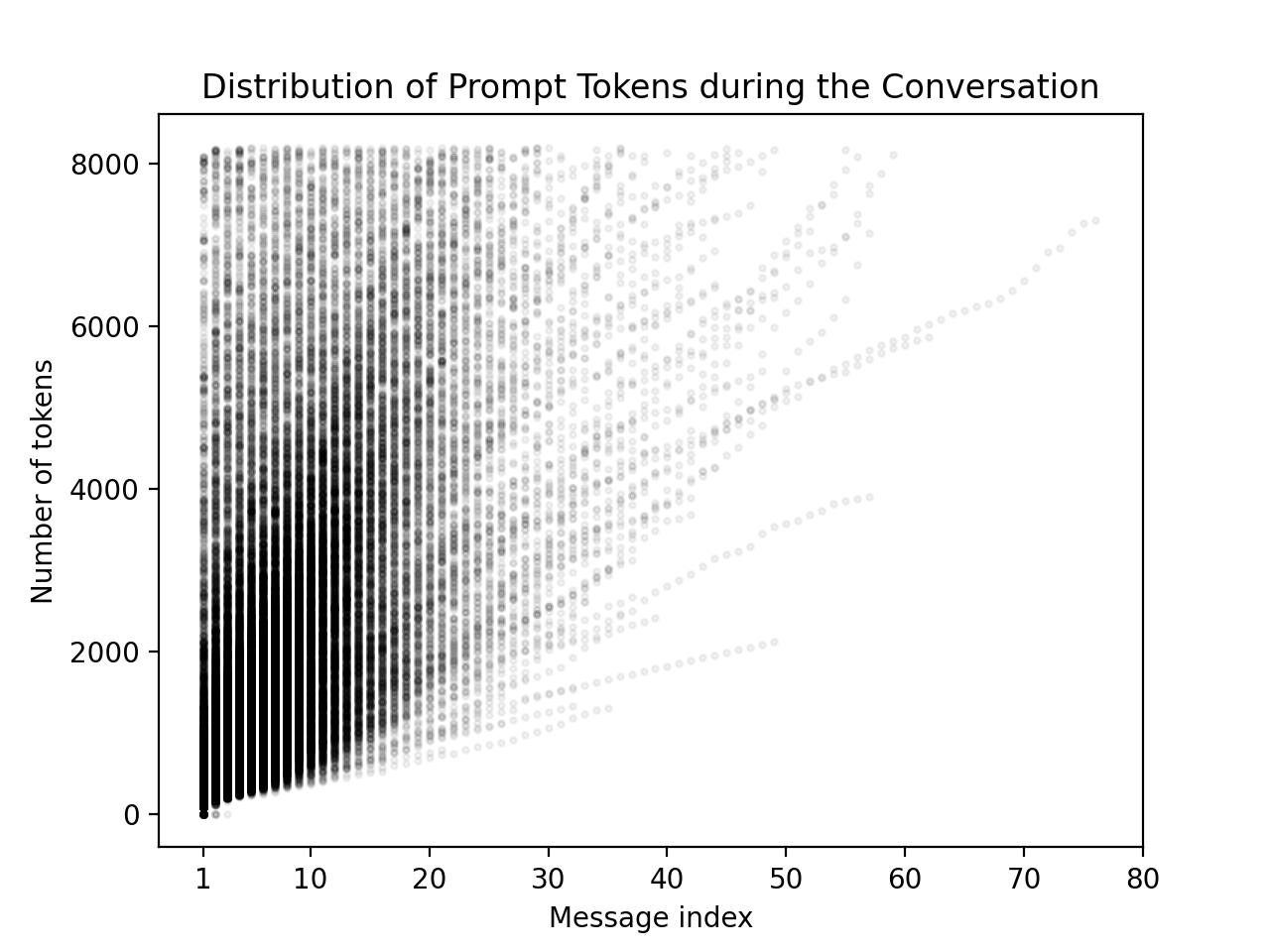}
    \caption{Distribution of prompt tokens during the conversation. A low alpha value is used for the dots to show the distribution of the data points.}
    \label{fig:distribution_of_prompt_tokens}
\end{figure}

\subsection{Data analysis}
Figure~\ref{fig:distribution_of_prompt_tokens} shows the distribution of the number of prompt tokens throughout the conversation. The data points are plotted using a low alpha value to visualize the concentration of the distribution. As expected, the number of tokens is increasing monotonically. It can be seen that most conversations start with a lower number of tokens and end after 15-20 interactions. However, there are also conversations that use the maximum number of tokens from the very first message, which in the current model is 8,192. Possibly, these conversations serve the purpose of generating text summaries or conducting code analyses. It is also noteworthy that, as indicated by certain outliers, a few conversations extend over a considerable number of messages.

\begin{figure}[ht]
\centering
\includegraphics[width=0.9\linewidth]{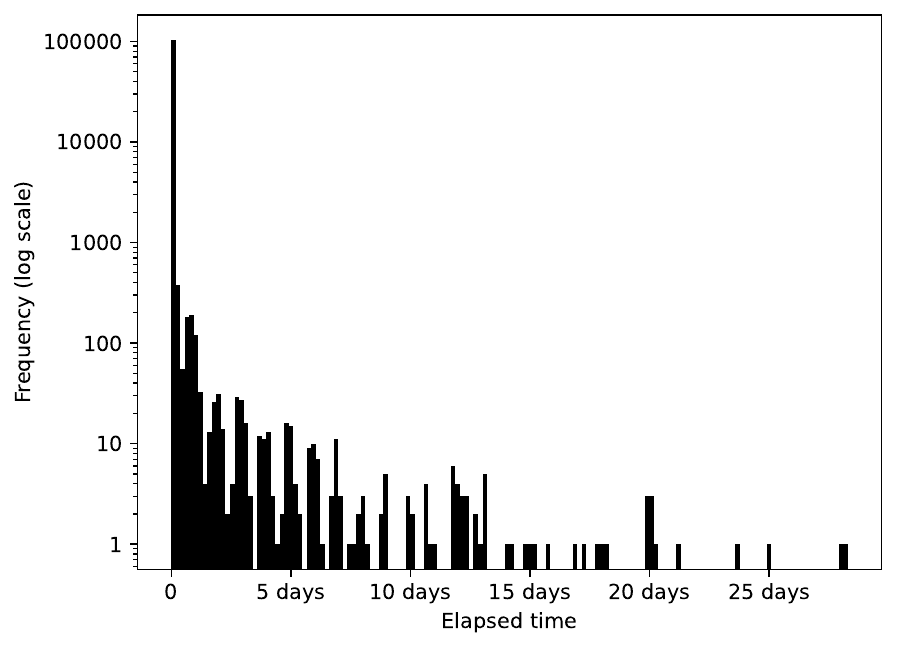}
    \caption{Length of pauses between messages. The y-axis is in logarithmic scale.}
    \label{fig:length_of_pauses}
\end{figure}

Next, the lengths of pauses between the messages of the same conversation were analyzed. Figure~\ref{fig:length_of_pauses} depicts these data in the form of a histogram. It is important to note that the y-axis is shown on a logarithmic scale. The vast majority of pauses are short, but the chart also shows that some pauses can last up to several days. 

\subsection{Simulation}
To carry out the study, it was necessary to detect shifts in context within the conversational threads. However, the data were anonymized and did not include any message content. Therefore, a proxy signal for these topic changes was established. Concretely, the duration of pauses between messages was used as an indicator, assuming that such breaks signified the beginning of a new conversation which made the previous messages irrelevant to the following ones and justified a reset of the conversational memory. To simulate this action, a new field \textit{new\_prompt\_tokens} was introduced, and whenever the pause exceeded a specific duration, the current tokens count for this and all following messages was reduced by the prompt tokens count of the previous record which supposedly contained outdated history.

\subsection{Results}

\begin{figure}[ht]
\centering
\includegraphics[width=0.9\linewidth]{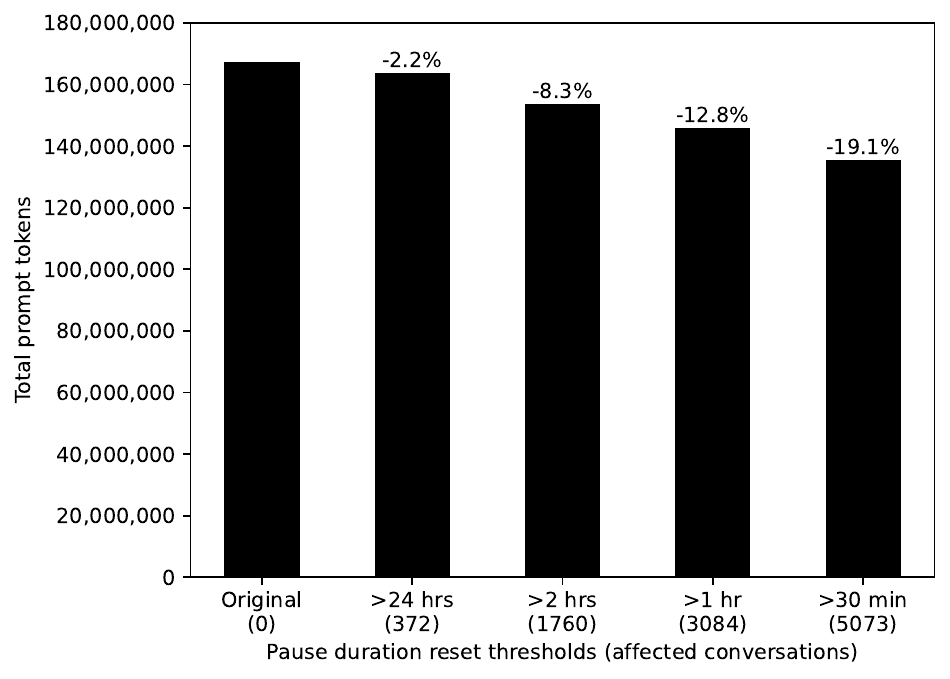}
    \caption{Simulated token saving potential. In the initial scenario (first bar), no tokens are discarded. In the subsequent scenarios, the conversational memory is reset when pauses of decreasing values are detected.}
    \label{fig:results}
\end{figure}

Figure~\ref{fig:results} depicts the results of simulating the implementation of the recommendation proposed in this paper. The y-axis displays the total number of prompt tokens used during the period under examination. The first bar of the chart represents the original scenario in which the conversational memory remains unaltered for any conversations, with some extending over 50 messages and lasting several days, as observed in Figures~\ref{fig:distribution_of_prompt_tokens} and \ref{fig:length_of_pauses}. The following bars in the chart demonstrate the impact of resetting the memory whenever the topic of the conversation has likely changed. Resetting the conversational memory after pauses of more than 24 hours affected 372 out of a total of 40,065 conversations, resulting in a 2.2\% reduction in prompt tokens. Employing shorter pauses as indicators for memory resets qualified more conversations and consequently yielded a higher reduction potential. For the shortest pause length tested in our simulation, pauses of 30 minutes or more, the conversational memory was reset in 5,073 conversation threads, leading to a 19.1\% reduction in prompt tokens.

As most cloud service providers which offer access to LLMs via API have a pay-per-token pricing model, the reduction can be directly translated into financial savings. Evaluating the environmental impact is less straightforward, however, due to the current lack of official emission reports for generative AI. As a solution, an estimate based on the measured energy consumption of a publicly available AI model is relied upon and then extrapolated to match the number of parameters in GPT-3.5-turbo. In the given case, an approximation suggests that the emission savings could range between 24.75 kg CO2e \footnote{\url{https://bit.ly/3UT0FEV}} and 69.33 kg CO2e\footnote{\url{https://bit.ly/3UUG03q}} if the recommended behavior were adopted.

\section{Conclusion}
This paper proposes as an effective eco gesture for the end users of conversational agents to reset the conversation whenever the topic or the purpose of the interaction changes and the preceding messages become irrelevant. Based on the observations above, such a measure can help reduce the number of tokens used, and thus the energy consumption, without compromising the quality of the answers.

For some users it may seem intuitive to interact in one long message thread with conversational agents due to their anthropomorphic characteristics that create the perception of a human-like dialog situation~\cite{ARAUJO2018183}. Considering the accumulation effect described here, a reassessment of this behavior could have a significant impact.

Identifying topic changes based on pauses is not entirely reliable. Thus, future work should include a study using data that also include the content of the messages. In addition, the investigation of more dynamic approaches to the automatic adaptation of the conversational memory should be considered: If abrupt changes of topic or longer idle times are detected, the system could reset itself to mitigate the observed adverse environmental impact.

\bibliographystyle{splncs04}
\bibliography{bibliography}

\begin{thebibliography}{10}
\providecommand{\url}[1]{\texttt{#1}}
\providecommand{\urlprefix}{URL }
\providecommand{\doi}[1]{https://doi.org/#1}

\bibitem{10.1162/coli_a_00474}
Apidianaki, M.: From word types to tokens and back: A survey of approaches to word meaning representation and interpretation. Computational Linguistics  \textbf{49}(2),  465--523 (06 2022). \doi{10.1162/coli_a_00474}, \url{https://doi.org/10.1162/coli{\_}a{\_}00474}

\bibitem{ARAUJO2018183}
Araujo, T.: Living up to the chatbot hype: The influence of anthropomorphic design cues and communicative agency framing on conversational agent and company perceptions. Computers in Human Behavior  \textbf{85},  183--189 (2018). \doi{https://doi.org/10.1016/j.chb.2018.03.051}, \url{https://www.sciencedirect.com/science/article/pii/S0747563218301560}

\bibitem{IPCC_2022_WGIII}
IPCC: Climate Change 2022: Mitigation of Climate Change. Contribution of Working Group III to the Sixth Assessment Report of the Intergovernmental Panel on Climate Change. Cambridge University Press, Cambridge, UK and New York, NY, USA (2022). \doi{10.1017/9781009157926}, \url{https://www.ipcc.ch/report/ar6/wg3/downloads/report/IPCC{\_}AR6{\_}WGIII{\_}FullReport.pdf}

\bibitem{kasneci2023chatgpt}
Kasneci, E., Se{\ss}ler, K., K{\"u}chemann, S., Bannert, M., Dementieva, D., Fischer, F., Gasser, U., Groh, G., G{\"u}nnemann, S., H{\"u}llermeier, E., et~al.: Chat{GPT} for good? on opportunities and challenges of large language models for education. Learning and individual differences  \textbf{103},  102274 (2023)

\bibitem{luccioni2023power}
Luccioni, A.S., Jernite, Y., Strubell, E.: Power hungry processing: Watts driving the cost of ai deployment? arXiv preprint arXiv:2311.16863  (2023)

\bibitem{luccioni2023estimating}
Luccioni, A.S., Viguier, S., Ligozat, A.L.: Estimating the carbon footprint of {BLOOM}, a 176b parameter language model. Journal of Machine Learning Research  \textbf{24}(253),  1--15 (2023)

\bibitem{Luccioni_2024}
Luccioni, S., Jernite, Y., Strubell, E.: Power hungry processing: Watts driving the cost of ai deployment? In: The 2024 ACM Conference on Fairness, Accountability, and Transparency. FAccT ’24, ACM (Jun 2024). \doi{10.1145/3630106.3658542}, \url{http://dx.doi.org/10.1145/3630106.3658542}

\bibitem{rapp2021human}
Rapp, A., Curti, L., Boldi, A.: The human side of human-chatbot interaction: A systematic literature review of ten years of research on text-based chatbots. International Journal of Human-Computer Studies  \textbf{151},  102630 (2021)

\bibitem{RAY2023121}
Ray, P.P.: Chat{GPT}: A comprehensive review on background, applications, key challenges, bias, ethics, limitations and future scope. Internet of Things and Cyber-Physical Systems  \textbf{3},  121--154 (2023). \doi{https://doi.org/10.1016/j.iotcps.2023.04.003}, \url{https://www.sciencedirect.com/science/article/pii/S266734522300024X}

\bibitem{rillig2023risks}
Rillig, M.C., {\AA}gerstrand, M., Bi, M., Gould, K.A., Sauerland, U.: Risks and benefits of large language models for the environment. Environmental Science \& Technology  \textbf{57}(9),  3464--3466 (2023)

\bibitem{rong2016optimizing}
Rong, H., Zhang, H., Xiao, S., Li, C., Hu, C.: Optimizing energy consumption for data centers. Renewable and Sustainable Energy Reviews  \textbf{58},  674--691 (2016)

\bibitem{10.1145/3581641.3584037}
Ross, S.I., Martinez, F., Houde, S., Muller, M., Weisz, J.D.: The programmer’s assistant: Conversational interaction with a large language model for software development. In: Proceedings of the 28th International Conference on Intelligent User Interfaces. p. 491–514. IUI '23, Association for Computing Machinery, New York, NY, USA (2023). \doi{10.1145/3581641.3584037}, \url{https://doi.org/10.1145/3581641.3584037}

\bibitem{10.1145/3381831}
Schwartz, R., Dodge, J., Smith, N.A., Etzioni, O.: {G}reen {AI}. Commun. ACM  \textbf{63}(12),  54–63 (Nov 2020). \doi{10.1145/3381831}, \url{https://doi.org/10.1145/3381831}

\bibitem{RePEc:eme:srjpps:srj-05-2020-0203}
Sheoran, M., Kumar, D.: {Benchmarking the barriers of sustainable consumer behaviour}. Social Responsibility Journal  \textbf{18}(1),  19--42 (December 2020). \doi{10.1108/SRJ-05-2020-0203}, \url{https://ideas.repec.org/a/eme/srjpps/srj-05-2020-0203.html}

\bibitem{thirunavukarasu2023large}
Thirunavukarasu, A.J., Ting, D.S.J., Elangovan, K., Gutierrez, L., Tan, T.F., Ting, D.S.W.: Large language models in medicine. Nature medicine  \textbf{29}(8),  1930--1940 (2023)

\bibitem{NIPS2017_3f5ee243}
Vaswani, A., Shazeer, N., Parmar, N., Uszkoreit, J., Jones, L., Gomez, A.N., Kaiser, {\L}., Polosukhin, I.: Attention is all you need. In: Guyon, I., Luxburg, U.V., Bengio, S., Wallach, H., Fergus, R., Vishwanathan, S., Garnett, R. (eds.) Advances in Neural Information Processing Systems. vol.~30. Curran Associates, Inc. (2017), \url{https://proceedings.neurips.cc/paper{\_}files/paper/2017/file/3f5ee243547dee91fbd053c1c4a845aa-Paper.pdf}

\bibitem{xu-etal-2022-beyond}
Xu, J., Szlam, A., Weston, J.: Beyond goldfish memory: Long-term open-domain conversation. In: Muresan, S., Nakov, P., Villavicencio, A. (eds.) Proceedings of the 60th Annual Meeting of the Association for Computational Linguistics (Volume 1: Long Papers). pp. 5180--5197. Association for Computational Linguistics, Dublin, Ireland (May 2022). \doi{10.18653/v1/2022.acl-long.356}, \url{https://aclanthology.org/2022.acl-long.356}

\bibitem{zhong-etal-2022-less}
Zhong, H., Dou, Z., Zhu, Y., Qian, H., Wen, J.R.: Less is more: Learning to refine dialogue history for personalized dialogue generation. In: Carpuat, M., de~Marneffe, M.C., Meza~Ruiz, I.V. (eds.) Proceedings of the 2022 Conference of the North American Chapter of the Association for Computational Linguistics: Human Language Technologies. pp. 5808--5820. Association for Computational Linguistics, Seattle, United States (Jul 2022). \doi{10.18653/v1/2022.naacl-main.426}, \url{https://aclanthology.org/2022.naacl-main.426}

\bibitem{zimmermann2021digital}
Zimmermann, S., Hein, A., Schulz, T., Gewald, H., Krcmar, H.: Digital nudging toward pro-environmental behavior: A literature review. PACIS p.~226 (2021)

\end{thebibliography}
\end{document}